\documentclass[usenatbib]{mn2e}
\usepackage{graphicx}

\begin{document}

\title[A Limit on $z=7$ LBGs]{A Limit on the Number Density of Bright $\mathbf{z\approx7}$ Galaxies}
\author[E.~R.~Stanway et al.]{Elizabeth R. Stanway\,$^{1}$, Malcolm N. Bremer\,$^{1}$, Valentina Squitieri\,$^{1}$,\newauthor 
Laura S. Douglas\,$^{1,2}$ \& Matthew D. Lehnert\,$^{3}$\\
$^{1}$\,H H Wills Physics Laboratory, Tyndall Avenue, Bristol, BS8 1TL, U.K.\\
$^{2}$\,School of Physics, Stocker Road, Exeter, EX4 4QL, U.K.\\
$^{3}$Laboratoire d'Etudes des Galaxies, Etoiles, Physique et Instrumentation GEPI, Observatoire de Paris, Meudon, France}

\date{Accepted 2008 January 28.  Received 2008 January 28; in original form 2007 October 4}

\pagerange{\pageref{firstpage}--\pageref{lastpage}} \pubyear{}

\maketitle

\label{firstpage}

\begin{abstract}
  
  We present a survey of bright optical dropout sources in two deep,
  multiwavelength surveys comprising eleven widely-separated fields,
  aimed at constraining the galaxy luminosity function at $z\approx7$
  for sources at 5-10L*($z=6$).  Our combined survey area is
  225\,arcmin$^2$ to a depth of $J_{AB}=24.2$ (3\,$\sigma$) and
  135\,arcmin$^2$ to $J=25.3$ (4\,$\sigma$).  We find that infrared
  data longwards of 2$\mu$m is essential for classifying optical
  dropout sources, and in particular for identifying cool Galactic
  star contaminants. Our limits on the number density of high redshift
  sources are consistent with current estimates of the Lyman break
  galaxy luminosity function at $z=6$.

\end{abstract}
\begin{keywords}
galaxies: high-redshift -- galaxies: starburst -- galaxies: evolution 
\end{keywords}


\section{Introduction}
\label{sec:intro}

Direct observations of galaxies at very high redshift can cast light
on the details of structure formation and early galaxy evolution. In
recent years, the most commonly observed tracer of star formation at
high redshift has been the Lyman Break galaxy (LBG) population
\citep{1999ApJ...519....1S}.  These sources are selected
photometrically based on a spectral feature interpreted as being the
$\lambda_\mathrm{rest}=$1215.67\AA\ Lyman break. Neutral
hydrogen along the line of sight leads to a forest of redshifted
Lyman-$\alpha$ absorption lines,
blue continuum flux is suppressed, and a
galaxy will exhibit red colours in a pair of filters bracketing the
break. Depending on the relative depth of the imaging in different
bands the galaxy will be seen to fall in magnitude between the red and
bluewards bands, or even to drop below the detection limit.  As a
result, Lyman break galaxies are often termed `drops' or `dropouts'
and their approximate redshift is determined by the band bluewards of
the break. Hence 
`U'- and `G'-drops lie at $z\approx3$ and $z\approx4$
\citep{1999ApJ...519....1S}, `V'-
and `R'-drops at $z\approx5$
\citep{2003ApJ...593..630L,2004ApJ...600L.103G} and `I'-drops at
$z\approx6$ \citep{2003MNRAS.342..439S,2006ApJ...653...53B}.

The use of LBG samples to study star formation at ever-increasing
redshifts has recently encountered a practical barrier.  At
$z\approx7$ the Lyman break is redshifted to 1\,$\mu$m, a spectral
region with poor sensitivity in both optical CCD detectors and
near-infrared instruments.
Photometry in the relatively broad photometric filters used longwards
of 1\,$\mu$m effectively smears out the signatures of sharp-sided
spectral features such as the Lyman break, making them difficult to
distinguish from more gradual spectral features such as a Balmer break
or molecular absorption in low redshift sources. The small field of
view, high background, low sensitivity and low multiplexing of current
near-infrared spectrographs further complicate the follow-up of Lyman
break galaxy candidates.

Sensitive, wide-format near-infrared imagers and multi-object
spectrographs on the largest telescopes offer potential for $z=7$
Lyman Break galaxy searches, identifying bright near-infrared sources
that drop in the optical bands (as discussed below in section
\ref{sec:col_sel}). Such surveys require knowledge of the $z=7$
luminosity function in order to estimate the required survey depth in
a given area. High contamination rates in optical-dropout samples also
complicate such surveys. While the most promising candidates will
always require spectroscopic follow-up, it is clearly necessary to
develop strategies for minimising the number of contaminating lower
redshift sources from photometry alone.  An important tool for
fulfilling these requirements is the {\em Spitzer Space Telescope},
which images at infrared wavelengths.  However the 85cm aperture of
{\em Spitzer} and 1$''.$2$\times$1$''.$2 arcsecond pixels of its IRAC
instrument lead to blending and confusion issues at faint limits
($m_{AB}>24$).  Existing deep field surveys incorporate sensitive
imaging at wavelengths from 4000\AA\ to 8$\mu$m (the four bands of the
IRAC instrument). As a result, they can provide a first constraint on
the surface density of bright galaxies at very high redshifts, and a
test of the ability of {\em Spitzer} to distinguish genuine $z>7$
sources from lower redshift contaminants.

In this paper we present an analysis of two multiwavelength
data-sets, identifying and classifying optical-dropout sources in
eleven widely separated fields, averaging across cosmic variance. In
section \ref{sec:col_sel} we discuss the colour selection criteria
which isolate high redshift galaxies. In sections \ref{sec:ergs} and
\ref{sec:goods} we apply such criteria to the ESO Remote Galaxy Survey
and the Great Observatories Origins Deep Survey respectively and
consider the resulting candidate sources. Finally, in section
\ref{sec:discussion} we discuss the implications of this analysis for
both the $z=7$ Lyman Break Galaxy luminosity function and future
wide-area near-infrared survey.

All magnitudes in this paper
 (optical and infrared) are quoted in the AB system 
 \citep{1983ApJ...266..713O}.  We use a flat 
 Universe with $\Omega_{\Lambda}=0.7$, $\Omega_{M}=0.3$ and $H_{0}=70
 h_{70} {\rm km\,s}^{-1}\,{\rm Mpc}^{-1}$.


\section{Colour Selection of $z=7$ Galaxies}
\label{sec:col_sel}

\begin{figure}
\includegraphics[width=8cm]{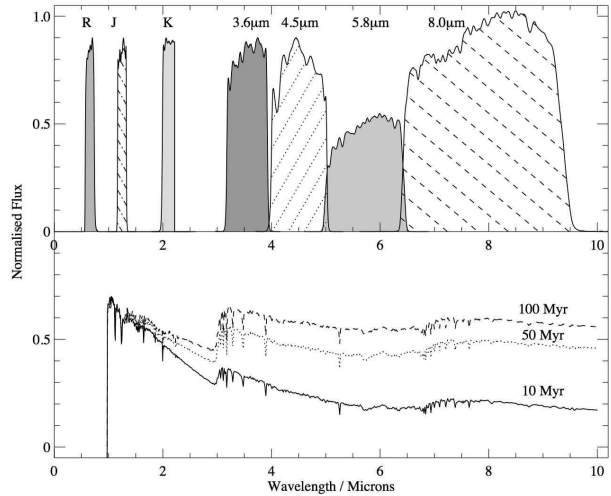}
\caption{The filters and model spectra considered in this work. The filter
  response curves of the ESO/VLT $R$-band, Mauna Kea Observatories $J$
  and $K$, and Spitzer/IRAC channels, span the rest-frame ultraviolet
  and optical spectra of galaxies at high redshift. We use the models
  of Maraston (2005), shown here placed at $z=7$.\label{fig:filters}}
\end{figure}

We employ the Lyman Break technique, identifying sources
that drop dramatically in flux between the reddest optical filter
($z$-band) and the bluest commonly used filter in the infrared
($J$-band). This criterion is sensitive to galaxies at
$z>7$, at which redshift flux longwards of the Lyman break becomes
shifted beyond the optical.  Given that the mean transmitted flux
fraction through the Lyman-$\alpha$ forest at $z\approx6$ is less than
5\% \citep{2002AJ....123.2183S} and may well fall to zero between
$z\approx6.0$ and $z\approx6.5$ \citep{2006AJ....132..117F}, genuine
high redshift candidates are expected to have no detectable flux in
any optical band, as figure \ref{fig:filters} illustrates.

\begin{figure}
\includegraphics[width=8cm]{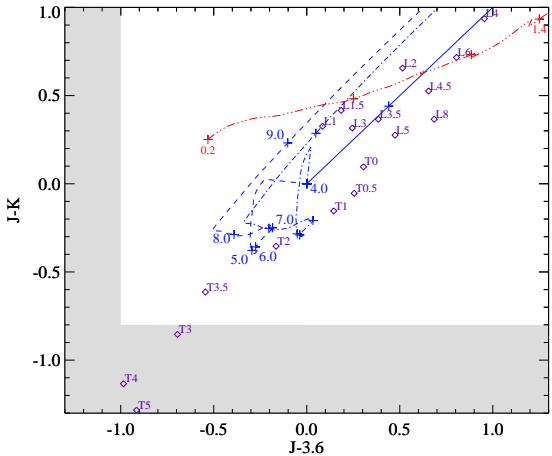}
\caption{ The near-infrared colours of optical dropout populations. 
  The blue lines show the redshift evolution of young starburst
  galaxies (solid -- flat in $f_\nu$; dashed -- 50\,Myr after start of
  burst, dot-dash -- 100\,Myr). The red line indicates the redshift
  evolution of an elliptical galaxy formed at $z=5$ with exponentially
  decaying star formation ($\tau$=0.5\,Gyr). The models of
  \citet{2005MNRAS.362..799M} are used throughout.  The purple
  diamonds indicate the measured colours of cool Galactic stars
  \citep{2006ApJ...651..502P}. Shaded regions are excluded by our
  colour selection criteria.
  \label{fig:jk_jc1}}
\end{figure}

\begin{figure}
\includegraphics[width=8cm]{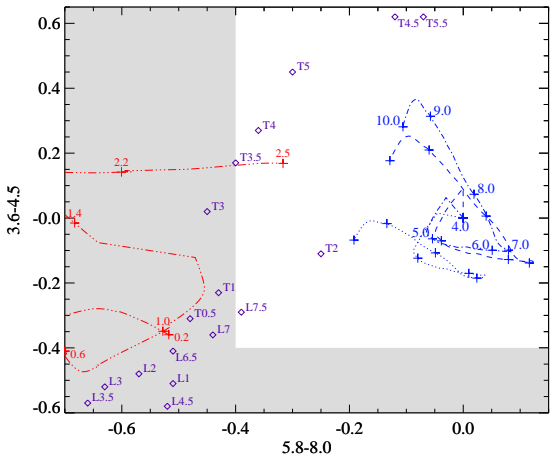}
\caption{The colours of optical dropout sources in the
  {\em Spitzer}/IRAC wavebands.
  For explanation of lines and symbols please see figure
  \ref{fig:jk_jc1}.
  \label{fig:j_c1c2}}
\end{figure}

However, other well known populations exhibit similar colours, with
relatively high number densities, rendering the search for high
redshift galaxies akin to the search for a needle in a haystack.  The
4000\AA\ Balmer break in old galaxies at $z\approx2$ could potentially
be mistaken for a Lyman break at $z>7$, although such galaxies are
likely to be detected in sufficiently deep optical imaging. Cool
late-type stars, particularly class L and T dwarfs, are also optically
faint, but potentially bright in the near-infrared
\citep[e.g.][]{2006ApJ...651..502P}.  In order to identify all or most
interlopers we require four broadband colour selection criteria in
addition to optical non-detection.  These are illustrated in figures
\ref{fig:jk_jc1} and \ref{fig:j_c1c2}.  Cuts in the $J-K$ or
$J-3.6\mu$m will effectively exclude the majority of cool $T$ class
stars which are blue in these colours. Since these two colours are
highly correlated, with most of the colour difference arising from
$J-K$, strictly only one is necessary. However, given the challenge of
observing in the thermal infrared from the ground, it is often
possible to survey large areas to deeper limits in 3.6$\mu$m with {\em
  Spitzer}/IRAC than in K from the ground, allowing the J-3.6$\mu$m colour
to provide stronger limits in the case of faint or undetected objects.

The true utility of the infrared, however lies at longer wavelengths.
As figure \ref{fig:j_c1c2} illustrates, a further requirement that
sources are flat or slightly red in the {\em Spitzer}/IRAC wavebands
is likely to exclude some low redshift galaxies and the majority of L
class stars, which separate neatly from the predicted colours of young
starburst galaxies at $z>7$.  Unfortunately, it is impossible to
exclude some late $L$ and early $T$ class stars while allowing for
photometric and intrinsic scatter in the high redshift galaxy
population.

A final source of contamination potentially arises from mature
($>$1\,Gyr old), highly reddened galaxies at intermediate redshifts.
Such sources have been identified in deep samples selected at
$3.6\mu$m and identified as optical dropouts
\citep[e.g.][]{2007A&A...470...21R,2006AJ....132.1405S} and have
extreme colours in both $J-K$ and $K-3.6\mu$m, as opposed to mature
galaxies at high redshift (which can have dramatic $J-3.6\mu$m
colours, but have near-zero colour in $J-K$. While it is impossible to
eliminate such sources from a sample without also rejecting
highly-reddened, mature galaxies (which show a Balmer break) at $z>8$,
any galaxies with such colours must be treated with caution.  Given
the scarcity of very high redshift galaxies, a Bayesian analysis of
the redshift likelihood distribution is likely to favour the low
redshift identification.

\section{The ESO Remote Galaxy Survey}
\label{sec:ergs}

 \subsection{Survey Outline}
 \label{sec:ergs_outline}
 
 The ESO Remote Galaxy Survey (ERGS, Douglas et al, in prep) is aimed
 at identifying $z>5$ Lyman break galaxies in the same deep imaging
 fields imaged by the ESO Distant Cluster Survey
 \citep[EDisCS,]{2005A&A...444..365W}. Each of ten, widely separated
 fields was surveyed in the optical ($VRIZ$) using the FORS2
 instrument on the Very Large Telescope (VLT), in the near-infrared
 ($JK_s$) using SOFI on the NTT, and in the infrared (3.6, 4.5, 5.8
 and 8.0$\mu$m) using IRAC on {\em Spitzer}. Typical depths reached
 were 27.8 in the $V$-band, 27.4 in $R$, 26.6 in $I$, 25.9 in $Z$,
 24.6 in the $J$-band and 23.8 in $K$ (2\,$\sigma$, 2$''$ apertures).
 Typical seeing was 0.6-0.8$''$ in the ground-based data.  In the IRAC
 bands, the 2\,$\sigma$ limiting depths were measured in a 4.5$''$
 aperture and were 25.1 at 3.6\,$\mu$m, 25.2 at 4.5\,$\mu$m, 23.0 at
 5.8\,$\mu$m and 22.9 at 8.0\,$\mu$m.  For full details of the data
 see \citet{2005A&A...444..365W} and Douglas et al (2008).  For the
 purposes of identifying optical dropouts, we required full ten band
 photometric coverage, yielding a total survey area of
 225.1\,arcmin$^2$ (after removing regions with high background noise
 in the near-infrared). Objects were selected in the $J$-band using
 the SExtractor software package \citep{1996A&AS..117..393B} and their
 fluxes measured at the centroid of the $J$-band detection in each
 band using the IDL `Aper' routine. For initial detection we required
 that an area of four contiguous pixels exceeded the background by at
 least 1.5 standard deviations, resulting in a basic photometry
 catalogue highly complete for unresolved and low surface brightness
 sources, but also highly contaminated by spurious objects at the
 faint end. Photometry was measured in 2$''$ apertures in the optical
 and near-infrared and in larger 4.5$''$ apertures in the {\em
   Spitzer}/IRAC bands in order to optimise signal-to-noise for our
 unresolved sources. The astrometry in the optical images and
 near-infrared images was found to be precise to better than 0.1$''$
 (half a pixel) and to within 0.6$''$ in the IRAC bands.  We find that
 adding an uncertainty of 0.35$''$ (half the seeing disk) to the
 location of the $J$-band centroid for our compact, low
 signal-to-noise objects has the effect of introducing an additional
 error to the photometry with a standard deviation of 0.04 magnitudes
 at the faint limit of our survey.  This is negligible when added in
 quadrature with the Poisson noise and statistical uncertainty in the
 background at the same magnitude.
 
 A refined catalogue was constructed of objects with $J$-band
 detections ($>$3\,$\sigma$), but no detection at 2\,$\sigma$ in each
 of the $V$, $R$, $I$ and $Z$ bands. The $V$, $R$ and $I$-band imaging
 was combined to create a single optical image (reaching a limiting
 magnitude better than 28.5 in each field) and a further non-detection
 required in this image. Magnitudes were not corrected for lensing.
 However, we note that the mean weak lensing effect across these
 fields is to brighten the observed magnitudes by 0.1\,magnitudes
 \citep{2006A&A...451..395C} compared to their intrinsic magnitudes.
 Candidate high redshift sources satisfy $J_{AB}<24.2$ (3-4$\sigma$
 depending on field) after correction for aperture losses.  Each
 candidate was inspected by eye and obvious data artifacts excluded,
 leaving a sample of 35 candidate sources for further inspection.
 
 By definition our target sources may only be bright in a single band,
 and this measurement is therefore susceptible to spurious sources in
 the detection band. In order to estimate the probability that our
 optical-dropout sources are not merely spurious detections of peaks
 in the random background noise, we repeated the analysis, inverting
 the J-band imaging in the ERGS fields and determining the number of
 negative noise peaks that would have been detected as objects to our
 limit if they were instead positive signals.
 
\begin{figure}
\includegraphics[width=8cm]{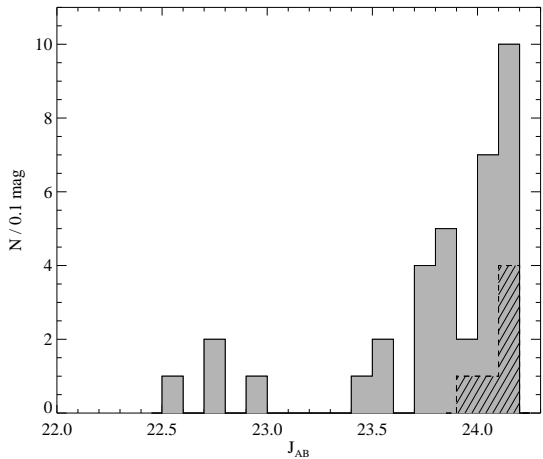}
\caption{The expected magnitude distribution of spurious sources due to fluctuations
  in background noise (hatched), compared with the distribution of
  optical-dropout candidates (shaded). The number of spurious sources is based on
  detection of the total number of negative noise peaks in the $J$-band imaging as
  a function of magnitude.
    \label{fig:inv_test}}
\end{figure}

 After discounting obviously spurious detections (cosmic ray
 subtraction residuals, regions of extended noise and depressions in
 the background in the wings of bright emission objects), we find six
 genuine noise peaks in the 225.1 arcmin$^2$ survey area which would
 have been mistaken for dropout sources if positive.  Their magnitude
 distribution, together with that of the optical dropout candidates is
 shown in figure \ref{fig:inv_test}.  Making the assumption that
 depressions in the background of this kind are statistically as
 likely as positive fluctuations which would reach the photometric
 catalogue, we estimate that a fraction of 17\% of our candidate sample,
 biased towards the faint end, may be spurious detections. We note
 that none of the spurious detections are brighter than `$J=23.9$'
 (assuming positive flux) and that none of them coincide with source
 detections in any of the other 9 bands surveyed.

Given the faint limits of our data, and the relatively broad PSF of
the {\em Spitzer Space Telescope} in the IRAC bands, we also test the
probability that random apertures placed on the 3.6$\mu$m images that
form part of the ERGS survey coincide with infrared-detected sources.
Placing 300 apertures randomly in each of the ten survey fields, we
find that 39\% of photometric apertures are expected to encounter flux
from contaminating sources equal to a 3\,$\sigma$ peak in the
background level (corresponding to a detection of $J=24.2$).

 \subsection{$z=7$ Candidates}
 \label{sec:ergs_selection}
 
 This analysis yielded a total of 35 optical dropout sources,
 distributed across 10 equally-sized fields. Of these the most densely
 populated field contained 7 candidate objects and the sparsest field
 only 1 candidate.
 
 The infrared colours of each candidate were subjected to detailed
 analysis to determine the most likely identification: $z>7$ galaxy,
 lower redshift galaxy or Galactic star.  Of the 35 dropout sources,
 15 were blended at the resolution of {\em Spitzer}/IRAC and cannot be
 studied further, consistent with the expected 13.7 confused sources
 given Poisson noise in the small number statistics. Deblending was
 not attempted since, at these faint magnitudes, errors in the
 resultant flux measurement would be dominated by uncertainty in the
 colour and isophotal profile of the confusing sources.  The remaining
 20 optical dropouts had clean infrared photometry. Of these, two
 sources were eliminated because of their blue $J-K$ colours
 ($J-K<$-0.8), and a further eight were rejected as too blue in the
 $J-3.6\mu$m colour ($J-3.6\mu$m$<-1.0$).  The most likely
 identification of these sources is that they are T-dwarf stars at
 moderate heliocentric distances (see section \ref{sec:tdwarfs}).
 Finally, a further nine candidates were eliminated from the sample as
 their colour (or the 2\,$\sigma$ limit on the colour) satisfied
 $3.6-4.5\mu$m$<-0.4$. These candidates are considered most likely to
 be either L-class dwarfs or old galaxies at intermediate redshift.
 
\begin{figure*}
\includegraphics[width=16cm]{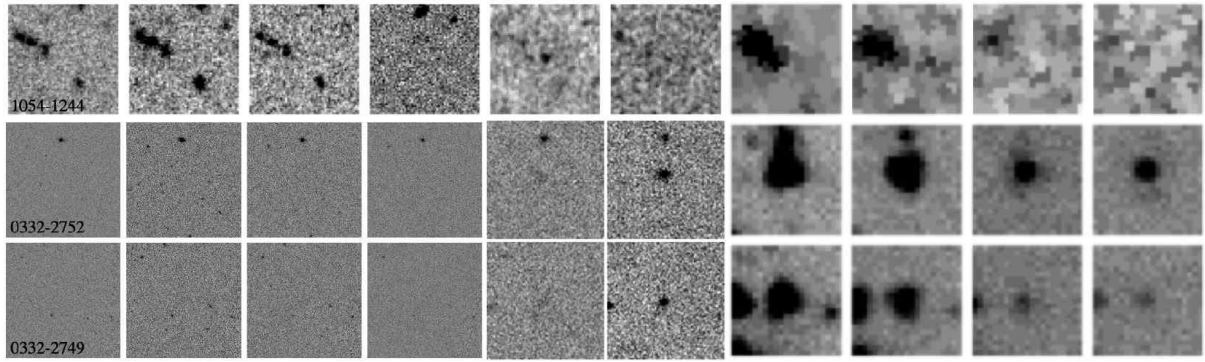}
\caption{Postage stamp images of the three candidate sources listed in table 
  \ref{tab:ergs_cand}. Boxes are 12 arcseconds to a side. In the case
  of 1054-1244 the bands shown are
  $V,R,I,Z,J,K,[3.6],[4.5],[5.8],[8.0]$. In the remaining two cases
  $V,R,I,Z$ is replaced by $b,v,i',z'$. While all three sources satisfy
the colour selection criteria, their behaviour in the infrared is quite different.
  \label{fig:stamps}}
\end{figure*}

Thus only one candidate of 20 sources studied has colours satisfying
the colour criteria in figures \ref{fig:jk_jc1} and \ref{fig:j_c1c2}.
The colours of this source are given in table \ref{tab:ergs_cand} and
it is shown in figure \ref{fig:stamps}. The remaining source has a
magnitude $J=23.6$ suggesting that it is statistically unlikely to be
a chance detection of a spurious noise peak. While the source
satisfies our formal selection criteria, its colours are close to the
bluewards cut-off in $3.6-4.5\mu$m colour. This colour is consistent
with that of an early T-dwarf and, given the photometric error bars on
this faint source, marginally consistent with identification as a
$z>7$ source in some star formation histories.  Hence, in our
225.1\,arcmin$^2$ survey area we find an upper limit of 1.43 sources,
assuming a similar contamination fraction for the infrared-blended
fraction, lying at $z>7$ and a more likely limit of fewer than 1 high
redshift galaxy to a depth of $J_{AB}=24.2$ (3$\sigma$).

  \begin{table*}
  \begin{tabular}{cccccccc}
  RA \& Declination (J2000) & $J_{AB}$           &  Opt$-J_{AB}$ &  $Z_{AB}-J_{AB}$  &    $J_{AB}-K_{AB}$      & ($J-3.6\mu$m)$_{AB}$        &  ($3.6-4.5\mu$m)$_{AB}$      &   ($5.8-8.0\mu$m)$_{AB}$ \\
  \hline\hline
10 54 41.35  -12 44 07.3  &  23.61$\pm$0.26 (3.8) &  $>$4.9  & $>$2.1  & $<$0.15       & -0.10$\pm$0.31     &  -0.36$\pm$0.27     &   -     \\
03 32 25.22  -27 52 30.7  &  24.73$\pm$0.19 (5.2) &  $>$3.3  & $>$1.8  & 1.81$\pm$0.22 &  3.35$\pm$0.20     &   0.45$\pm$0.06     &   -0.14$\pm$0.24\\
03 32 16.81  -27 49 07.5  &  24.79$\pm$0.22 (4.5) &  $>$3.2  & $>$2.1  & 1.57$\pm$0.25 &  2.99$\pm$0.23     &   0.05$\pm$0.06     &   -0.36$\pm$0.24\\
  \end{tabular}
\caption{Photometry of the remaining sources satisfying the colour selection
 criteria for high redshift ($z>7$) galaxies in the ERGS and GOODS survey areas. The Opt$-J$
 indicates an optical to near-IR colour, measured from the VRI combined image for the first
  object, and from the F435W band in the latter two candidates. The signal to noise of each
source in the $J$-band is shown in brackets.\label{tab:ergs_cand}}
\end{table*}

\section{The GOODS Survey}
\label{sec:goods}

 \subsection{Survey Outline}
 \label{sec:goods_outline}
 
 The Great Observatories Origins Deep Survey
 \citep[GOODS,][]{2004ApJ...600L..93G} is a publicly-available dataset
 comprising observations from major ground and space-based
 observatories. We make use of the r1.1 photometric catalogue release
 which comprises the results of imaging in the F404W ($b$), F606W
 ($v$), F814W ($i'$) and F850LP ($z'$) filters of the Advanced Camera
 for Surveys on the {\em Hubble Space Telescope} ({\em HST}/ACS). We
 supplement this with the $J$- and $K_s$-band data obtained as part of
 the GOODS programme at the VLT \citep{2006AA...454..423V} and with
 {\em Spitzer}/IRAC data obtained as part of the GOODS Legacy
 Programme (PI: Mark Dickinson). As in the analysis in section
 \ref{sec:ergs} we measured photometry in 2$''$ apertures in the
 optical and near-IR and in 4.5$''$ apertures in the infrared. In
 these large apertures we require non-detection to limiting magnitudes
 of 27.8 in $b$ and $v$, 27.0 in $i'$ and 26.8 in the $z'$-band and
 also non-detection in visual inspection of the images (since compact
 sources can be detected to more sensitive limits in the high
 resolution space-based data). The near infrared imaging reaches
 $2\,\sigma$ limits of $J=25.8$ and $K_s=25.2$, while the 2\,$\sigma$
 limiting depths were 27.1 at 3.6\,$\mu$m, 26.6 at 4.5\,$\mu$m, 24.8
 at 5.8\,$\mu$m and 8.0\,$\mu$m.
 
 We restrict our analysis to the GOODS-South field and survey a total
 of 135.0 arcmin$^2$ with full ten-band photometric coverage. As
 before, we train our catalogue in the $J$-band and measure the flux
 at the near-infrared location in $K_s$ and longwards. Optical
 counterparts are assigned to sources if an {\em HST}/ACS detection is
 identified within 1$''$ of the near-IR centroid. Optical dropout
 candidates were inspected by eye and spurious detections removed. We
 limit our catalogue to sources with a 4\,$\sigma$ detection of
 $J<25.3$.
 
 As in the case of the ERGS survey fields above, we consider the
 fraction of confused sources we might expect in the {\em
   Spitzer}/IRAC bands through the investigation of randomly placed
 apertures on the $3.6\mu$m image. While the IRAC data in the GOODS
 fields is deeper than in the ERGS fields, the GOODS region was chosen
 for its low foreground contamination, while the ERGS survey exploits
 lower redshift cluster fields. Hence the contamination fraction
 amongst randomly placed apertures in the GOODS-S field was a slightly
 lower fraction of 28\% coincidence with $3\sigma$ flux from other
 sources.

 \subsection{$z=7$ Candidates}
 \label{sec:goods_selection}
 
 This analysis yielded a total of 14 optical dropout sources.  Of
 these, 4 sources (1 with $J<24.2$) are blended at the resolution of
 {\em Spitzer}/IRAC and cannot be studied further, consistent with the
 3.9 sources expected from pure confusion.  A further 2 sources are
 too blue in their infrared colours to be high redshift galaxies (i.e.
 J-3.6$\mu$m$<-1.0$), and are tentatively classified as likely T-dwarf
 stars.  The remaining 8 sources are well-detected at 3.6-8.0$\mu$m.
 However, while their 3.6-4.5$\mu$m colours are consistent with either
 high or low redshift identification, in six cases (2 of which have
 $J<24.2$) their colours at longer wavelengths are rather blue
 (5.8-8.0$\mu$m$<-0.4$) suggesting that a low redshift interloper
 galaxy interpretation is more likely.
 
 The photometry of the remaining two candidates is given in table
 \ref{tab:ergs_cand}. We note that both sources are detected in the
 $J$-band at $>5$ sigma, suggesting that they are statistically
 unlikely to be spurious detections based on our analysis of the ERGS
 fields discussed above. Both sources are also well detected in the
 $K$-band, making a chance coincidence of $J$-band noise with a {\em
   Spitzer}/IRAC source unlikely. Both sources show a red spectrum
 leading to near equal $J-K$ and $K-3.6\mu$m colours. As discussed in
 section \ref{sec:col_sel}, this may suggest either identification as
 a high redshift candidate or as a reddened mature galaxy at
 intermediate redshifts. Given the scarcity of high redshift galaxies,
 the latter explanation is the more likely.  Hence, in our survey area
 of 135\,arcmin$^2$ we find an upper limit of 2.8 sources (accounting
 for the fraction of infrared-blended sources) lying at $z>7$ and a
 more likely limit of fewer than 1 high redshift galaxy to a depth of
 $J_{AB}=25.3$ (4$\sigma$).


\section{Discussion}
\label{sec:discussion}

 \subsection{The Surface Density of L and T-dwarf Stars}
 \label{sec:tdwarfs}
 
 Given the difficulty in unambiguously classifying extremely red
 sources on the basis of photometry alone, it is instructive to
 determine whether the red population identified in this work
 represents an excess over predicted number counts for low redshift
 objects.  Specifically, given that early- to mid-T dwarfs are
 expected to be our most significant contaminant on the basis of their
 observed infrared colours \citep{2006ApJ...651..502P}, it is useful
 to calculate the surface density of T-dwarf stars\footnote{We do not
   discuss the density of L dwarfs since an essentially unknown
   fraction, varying with distance, of these will be excluded by the
   optical non-detection criterion.}.

 The number of T-dwarf stars with spectroscopic confirmation in the
 literature remains small \citep[see][and references
 therein]{2007AJ....134.1162L}, although this number is increasing
 rapidly due to follow-up of photometrically-identified candidates in
 wide area surveys such as the SDSS
 \citep[e.g.][]{2006AJ....131.2722C}. \citet{2007AJ....134.1162L}
 report a lower limit to the surface density of T-dwarfs (T1-T8) of
 $3\times10^{-3}$\,pc$^{-3}$ based on a catalogue of known sources
 within 10\,pc of the sun , although they note that they may be
 considerably incomplete due to the non-uniformity of the input
 surveys. A more complete analysis of the surface density of T-dwarfs
 was performed in the SDSS DR1/2MASS overlap region by
 \citet{2007arXiv0710.4157M}, using Monte Carlo simulation to estimate
 the effect of selection functions in both surveys. They estimate a
 volume density of class T0-T8 dwarfs of
 $(7.0\pm3.1)\times10^{-3}$\,pc$^{-3}$.
 
 Given the typical absolute magnitude of a class T3 star
 \citep[$J_{AB}=15.3$,][]{2007arXiv0710.4157M}, we would be able to
 detect such a star between a distance of 57\,pc (closer than which
 our detectors saturate) and 603\,pc (for a limit of $J=24.2$) or
 1000\,pc (for $J=25.3$). Given that these distances are well within
 the scale height of the disk, we make the assumption that the volume
 density of T stars is constant in this range. From this, we estimate
 a total enclosed volume for early- to mid-T stars of 2224\,pc$^{3}$
 in our shallow survey and 3796\,pc$^{3}$ in our deep survey area.
 This corresponds to a predicted 16$\pm$7 T-dwarf stars in our shallow
 survey and 27$\pm$12 T stars in our deep survey.
 
 We compare this with the number of sources in our dataset classified
 as likely T-dwarf interlopers on the basis of their infrared colours.
 To a limit of $J=24.2$ in the eleven survey fields (ERGS+GOODS) we
 identified a total of ten sources as likely T-dwarf interlopers due
 to very blue colours in the {\em Spitzer}/IRAC bands. Assuming an
 equal ratio of T-dwarf candidates to $z$-drops amongst the
 infrared-confused sources, this yields a tentative volume density of
 $(7.9\pm0.2)\times10^{-3}$\,pc$^{-3}$.

We note that this is at the high end of the volume density estimates
discussed above, but is consistent within the error bars, given that
this is a near-infrared selected sample.  Given the huge field to
field variation in the number density of red sources in our shallow
survey, and the still significant uncertainty in the surface density
of local T dwarfs, our results are consistent with at least half of
our identified $z$-drops being T-dwarf stars.

Our results from the deep survey area are less consistent with
published volume densities for interlopers, with just two
optical-dropout sources identified as good T-dwarf candidates to a
depth of $J=25.3$. We note that our deep survey is confined to a
single region, the GOODS-S field, which also appears to be deficient
in $z$-band dropout sources at $J<24.2$ compared to the mean of our
ERGS survey fields, and that there is a factor of 6 in surface density
between our richest and sparsest fields. It is possible that the
GOODS-S sightline probes a void in the local T-dwarf distribution, or
that the local density is a poor approximation at increasing
heliocentric (and, in this case, galactocentric) distance.  Given the
very small solid angle subtended by our fields, it is likely that
multiple sightlines through the Galactic disk and halo will be
required to determine the mean mean volume density of such sources as
a function of Galactocentric coordinate.

 \subsection{Implications for the $z=7$ Luminosity Function}
 \label{sec:lf}
 
 While the detection of zero or very few dropout candidates to these
 relatively shallow limits is not entirely surprising, it is
 interesting to consider possible constraints on the luminosity
 function that may be derived from this result.
 
 The typical luminosity of the Lyman break galaxy population, M*, is
 known to evolve with cosmic time, becoming fainter with increasing
 redshift \citep{2007arXiv0707.2080B}. The highest redshift sample for
 which the luminosity function has been well constrained is the
 $i'$-dropout population at $z=6$. \citet{2007arXiv0707.2080B} found
 that such a population is described by a Schecter function with
 M*$_{UV}$=-20.2$\pm$0.2, $\alpha=-1.7$ and a surface density
 $\phi*=(1.4\pm0.5)\times10^{-3}$\,Mpc$^{-3}$.

Given no evolution in the shape and comoving volume density
normalisation luminosity function between $z=6$ and $z=7$ (914\,Myr
and 748\,Myr after the Big Bang), we can predict the expected number
of sources at the bright end of the luminosity function. Our colour
selection (described in section \ref{sec:col_sel}) is, in theory
sensitive over a broad redshift range, from $z=6.5$ to $z=10$.
However, in an apparent magnitude-limited sample, the redshift
sensitivity function will be biased towards sources at lower redshift,
particularly for a luminosity function with a steep faint-end. Our
limiting magnitude is 1.6\,mag fainter at $z=10$ than at $z=6.5$, a
dramatic difference at $>$L*. We therefore calculate our survey volume
using the effective volume method after \citet{1999ApJ...519....1S},
accounting for the different limiting luminosity in each $\Delta
z=0.01$ redshift shell, using the prescription
\[
V_\mathrm{eff}(m)=\int dz\,p(m,z)\,\frac{dV}{dz}
\]
where $p(m,z)$ is the probability of detecting a galaxy at redshift
$z$ and apparent $z$-band magnitude $m$, given the luminosity function and
galaxy colour (randomly perturbed by an error of 0.3 magnitudes at the
faint limit of the survey), and $dz\,\frac{dV}{dz}$ is the comoving
volume per unit solid angle in a slice $dz$ at redshift $z$.

Given the parameters of the $z=5.9$ luminosity function discussed
above \citep{2007arXiv0707.2080B}, we find that our shallow survey is
sensitive to 8.0\% of the total volume enclosed between $z=6.5$ and
$z=10.0$ per unit area to the limiting luminosity of the faintest
object in the sample, while the deep survey selection is sensitive to
16.5\% of the same volume. We note that if we were to adopt instead
the tentative value of M* proposed by \citet{2007arXiv0707.2080B} for
$z=7$, we would survey an effective volume only 4.5\% and 9\% of the
total volume enclosed by the redshift range sensitivity range.

In our shallow sample reaching $J=24.2$, we survey an effective volume of
$1.0\times10^6$\,Mpc$^3$ and reach a typical limiting magnitude (at
$z=7$) of $M_{UV}$=-22.7.  Assuming no evolution in the high redshift
LBG luminosity function, we are sensitive to galaxies brighter than
10\,M* (given the z=6 value for M*). Our detection of no good
candidates in the combined GOODS+ERGS sample is consistent with the
predicted 0.001 galaxies.

While this is not unexpected it is interesting to note that
recent evidence from $z>5$ galaxy populations have suggested that the
the universe at earlier times was a more vigorously active place.  The
evolution in size of Lyman-break galaxies, together with their more
slowly evolving star formation rates implies that star formation
density rises with increasing redshift \citep{2007MNRAS.377.1024V}.
The identification of old stellar populations in $z=6$ galaxies
\citep{2005MNRAS.364..443E} also implies that starburst activity must
have been rapid and intense at early times. Hypothetical luminosity
functions for primordial starbursts (which have a
top-heavy initial mass function and a higher light to mass ratio than
more conventional stellar populations) imply that a
factor of 100 or more increase might be seen in the number density of
ultraviolet-luminous sources at $z>7$ \citep{2004ApJ...604L...1B}.
 The number density of $z=7$
candidates in our combined ERGS+GOODS survey area provides a firm upper
limit on the bright end of the luminosity function, confirming that it
inhabits the expected region of parameter space. 

Our deeper sample reaches $J=25.3$ or a typical limiting magnitude (at
$z=7$) of $M_{UV}$=-21.6. Assuming no evolution in the high redshift
LBG luminosity function, we are sensitive to galaxies brighter than
3.6\,M*. In our effective survey volume, $7.7\times10^5$\,Mpc$^3$, we
might expect to observe 2.4 galaxies, consistent with our detection of
up to two good candidates. This is sufficient to rule out
the most optimistic predictions for metal-free stellar populations at
$z>7$ \citep[see][]{2004ApJ...604L...1B}.  The upper limits derived
from our analysis are consistent with the measured LBG luminosity
function at $z=6$, and suggest that the volume density of luminous
(brighter than M*) galaxies remains approximately constant (if both
our candidates are real $z>7$ galaxies) or decreases (if one or more
is not real) with increasing redshift rather than increasing.

\citet{2004ApJ...616L..79B} estimated the surface density of optical
dropout sources in the Hubble Ultra Deep Field and identified four
likely candidates to a depth of $H_{AB}=27.0$ (as opposed to 8
predicted $z=7$ galaxies based on the $i'$-drop luminosity function).
Given a spectrum flat in $f_\nu$ (as appropriate for young
starbursts), this is equivalent to two magnitudes deeper than our
limit in the GOODS fields and the area of their survey was only 4\% of
that discussed here. This would imply an expected 0.5 $z=7$ galaxies
meeting our deep survey criteria. While we determine an upper limit of
2 galaxies at $z>7$ in our deep survey, our more realistic expectation
is that neither is likely to lie at high redshift. The confirmation of
one or zero sources would confirm that the HUDF is not significantly
under-abundant compared with other sightlines at the same redshift.

Given the varied colour selections employed by different authors, fair
comparisons are difficult.  We note that we would have rejected the
majority of massive $z>4$ galaxy candidates proposed in the
GOODS-South by \citet{2007MNRAS.376.1054D} due to the presence of flux
in the observed-frame optical or as below the reliable detection limit
in the $J$-band. By contrast the requirement of
\citet{2007A&A...470...21R} that their candidates be fainter than
$K_s=23.5$ would have excluded a number of the candidate galaxies
presented here. Their selection method, based on a flux-limited
3.6$\mu$m sample, may effectively select massive galaxies at these
redshifts, and yet is likely to suffer incompleteness due to both
confusion and insensitivity to young, starbursts \citep[which are
highly stochastic and dominate the number counts in high redshift
Lyman break galaxy samples][]{2007MNRAS.377.1024V}.  Hence the
optical-dropout survey presented here is complementary to other $z>7$
surveys.

 \subsection{Implications for Future Surveys}
 \label{sec:implications}
 
 In the next few years, the UKIDSS Ultra Deep Survey (UDS) will survey
 an area of 0.8\,deg$^2$ to $K_{AB}\approx25$ and will
 incorporate a search for high redshift galaxies. Our deep survey
 probes 12.5\% of this area and find at most 2 candidates which might
 require spectroscopic follow-up, consistant with 16
 potential $z=7$ candidates in the full volume of the UDS - a feasible
 number given a reasonable investment of large telescope time.  This
 modest number of galaxies is largely independent of evolution in the
 faint-end slope of the luminosity function, since the selection is
 truncated well above M*. Given the $z=6$ luminosity function, perhaps
 seven of the candidate galaxies identified in the UDS might be
 expected to lie at high redshift, although continued evolution in M*
 with increasing redshift potentially reduces this number still
 further.  It is clear that, even with exceptionally deep auxiliary
 data bluewards and redwards of the selection bands, the ratio of
 candidates to genuine galaxies in a photometric sample is likely to
 be at least 2.3:1.
 
 However, as the analysis in this paper makes clear, the number of
 contaminant sources in samples derived from optical/near-infrared
 photometry alone is likely to be much higher. Depth-matched imaging
 in the infrared wavebands of {\em Spitzer}/IRAC are essential for
 reducing the number of near-infrared photometric candidates (which in
 the UKIDSS UDS could exceed a hundred) to a more practical number for
 expensive follow-up observations.
 
 Wide-field infrared imagers currently being commissioned, such as
 Hawk-I on the VLT, are likely to revolutionise this field, making
 $z\ge7$ as accessible to observers as $z<6$ at the current time.
 However, the fundamental requirement that auxiliary observations at
 longer wavelengths are obtained above the atmosphere will limit their
 utility for high-redshift surveys after the inevitable decline of
 IRAC on the {\em Spitzer Space Telescope} and until its eventual
 replacement by the {\em James Webb Space Telescope}.

One prospect for the hiatus period between these two events is to
extend the lifetime of {\em Spitzer} in a `warm' mode
\citep{2007arXiv0709.0946V}.  While sensitivity would be lost
longwards of 5\,microns, the two shortest {\em Spitzer}/IRAC wavebands
provide sufficient spectral leverage to identify the majority of
potential contaminants ($>$80\% in this study), and in particular to
isolate cool Galactic stars from the galaxy locus. Inevitably the size
of the `haystack'
will increase with the loss of the 5.8 and 8.0$\mu$m wavebands.
However, the use of a warm {\em Spitzer} for surveys would
provide an archive that would be of use for many years,
complementing both existing and future, deep ground-based
surveys.


\section{Conclusions}
\label{sec:conc}

The key points presented in this paper can be summarised as follows:

i) We have identified an analysed a sample of $z$-band dropout sources
in deep, multi-wavelength imaging, considering their suitability as
potential $z>7$ galaxy candidates.

ii) In an area of 360\,arcmin$^2$ (the ERGS+GOODS surveys), we determine
 a strict upper limit of 1.4 potential $z=7$ galaxies and a likely limit
 of zero candidates to $J=24.2$. 

iii) In an area of 135\,arcmin$^2$ (the GOODS survey), we determine
 a strict upper limit of 2.8 potential $z=7$ galaxies and a likely limit
 of no good candidates to $J=25.3$.

iv) These results are consistent with current estimates of the luminosity
function of Lyman break galaxies at $z=6$, for galaxies with $L>5L*$.

v) The use of deep data longward of the $K$-band is extremely valuable if the
number of high redshift candidates from deep surveys such as UKIDSS
UDS from unmanageable to reasonable numbers for spectroscopic
follow-up.


\subsection*{Acknowledgements}

ERS gratefully acknowledges support from the UK Science and Technology
Facilities Council (STFC).  Based in part on observations made with
the NASA/ESA Hubble Space Telescope, obtained from the Data Archive at
the Space Telescope Science Institute. Also based in part on
observations made with the Spitzer Space Telescope, which is operated
by the Jet Propulsion Laboratory, CalTech. We thank the GOODS team for
making these high quality datasets public. Research presented here has
benefitted from the M, L, and T dwarf compendium housed at
DwarfArchives.org and maintained by Chris Gelino, Davy Kirkpatrick,
and Adam Burgasser. Results from the EDisCS/ERGS fields are based on
observations made with ESO telescopes under programmes 166.A-0162 and
175.A-0706.

\label{lastpage}

\end{document}